\documentclass[11pt]{article}
\usepackage{cite}
\usepackage{epsfig}
\usepackage{rotating}
\usepackage{url}
\usepackage{amsmath}
\usepackage{amsfonts}   
\usepackage{graphics}  
\usepackage{psfrag} 
\usepackage{graphicx} 
\usepackage[latin1]{inputenc}
\usepackage{color}
\usepackage{dcolumn}
\usepackage{bm}

\newcommand{\be}{\begin{equation}}
\newcommand{\ee}{\end{equation}}
\newcommand{\ba}{\begin{eqnarray}}
\newcommand{\ea}{\end{eqnarray}}

\newcommand \gtwid {{\mathrel{\raise.3ex\hbox{$>$\kern-.75em\lower1ex\hbox{$\sim$}}}}}
\newcommand \ltwid {{\mathrel{\raise.3ex\hbox{$<$\kern-.75em\lower1ex\hbox{$\sim$}}}}}
\newcommand \CE {{\mathcal{E}}}
\newcommand \CP {{\mathcal{P}}}

\begin{document}

\begin{center}
  \Large{\textbf{Fractal and Multifractal Scaling of Electrical Conduction in Random Resistor Networks}}\\\smallskip
  \normalsize{S. Redner}
\end{center}
\begin{center}
\normalsize{
Center for Polymer Studies and Department of Physics, Boston University\\ 
590 Commonwealth Ave., Boston, MA~ 02215~ USA
}
\end{center}

\vskip -0.2in

\noindent 
\section*{Article Outline}
Glossary\\
I.       Definition of the Subject\\
II.      Introduction to Current Flows\\
III.      Solving Resistor Networks\\
\indent III.1 Fourier Transform \\
\indent III.2 Direct Matrix Solution\\
\indent III.3 Potts Model Connection\\
\indent III.4 $\Delta$-Y and Y-$\Delta$ Transforms\\
\indent III.5 Effective Medium Theory\\
IV. Conduction Near the Percolation Threshold\\
\indent IV.1 Scaling Behavior \\
\indent IV.2 Conductance Exponent\\
V. Voltage Distribution of Random Resistor Networks\\
\indent V.1 Multifractal Scaling \\
\indent V.2 Maximum Voltage\\
VI.  Random Walks and Resistor Networks\\
\indent VI.1 The Basic Relation\\
\indent VI.2 Network Resistance  and P\'olya's Theorem\\
VII. Future Directions\\
VIII. Bibliography

\section*{Glossary}

{\small

  \noindent {\bf conductance ($G$)}: the relation between the current $I$ in
  an electrical network and the applied voltage $V$: $I=GV$.  \vskip 0.10cm

  \noindent {\bf conductance exponent ($t$)}: the relation between the
  conductance $G$ and the resistor (or conductor) concentration $p$ near the
  percolation threshold: $G\sim (p-p_c)^t$.  \vskip 0.10cm

  \noindent {\bf effective medium theory (EMT)}: a theory to calculate the
  conductance of a heterogeneous system that is based on a homogenization
  procedure.  \vskip 0.10cm

  \noindent{\bf fractal}: a geometrical object that is invariant at any
  scale of magnification or reduction. \vskip 0.10cm

  \noindent{\bf multifractal}: a generalization of a fractal in which
  different subsets of an object have different scaling behaviors.\vskip 0.10cm

  \noindent {\bf percolation}: connectivity of a random porous network.\vskip
  0.10cm

  \noindent {\bf percolation threshold $p_c$}: the transition between a
  connected and disconnected network as the density of links is varied.\vskip
  0.10cm

  \noindent {\bf random resistor network}: a percolation network in which the
  connections consist of electrical resistors that are present with
  probability $p$ and absent with probability $1-p$.

}

\section{Definition of the Subject}
\label{sec:def}

\begin{figure}[ht]
  \begin{center}
    \includegraphics[width=0.65\textwidth]{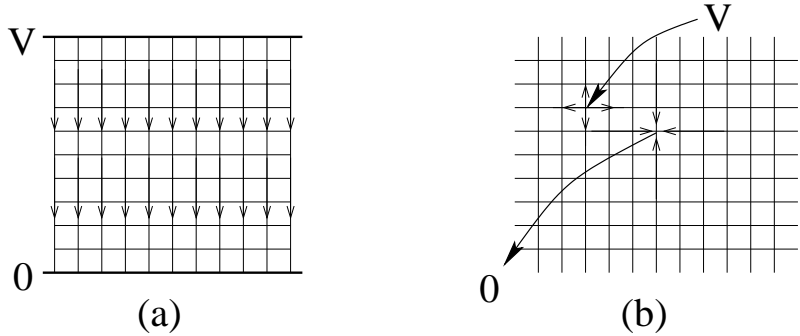}
    \caption{Resistor networks in the (a) bus-bar geometry, and (b) the
      two-point geometry.}
\label{def}
  \end{center}
\end{figure}

Consider an arbitrary network of nodes connected by links, each of which is a
resistor with a specified electrical resistance.  Suppose that this network
is connected to the leads of a battery.  Two natural scenarios are: (a) the
``bus-bar geometry'' (Fig.~\ref{def}), in which the network is connected to
two parallel lines (in two dimensions), plates (in three dimensions), {\it
  etc.}, and the battery is connected across the two plates, and (b) the
``two-point geometry'', in which a battery is connected to two distinct
nodes, so that a current $I$ injected at a one node and the same current
withdrawn from the other node.  In both cases, a basic question is: what is
the nature of the current flow through the network?

There are many reasons why current flows in resistor networks have been the
focus of more than a century of research.  First, understanding currents in
networks is one of the earliest subjects in electrical engineering.  Second,
the development of this topic has been characterized by beautiful
mathematical advancements, such as Kirchhoff's formal solution for current
flows in networks in terms of tree matrices \cite{K47}, symmetry arguments to
determine the electrical conductance of continuous two-component media
\cite{R92,B35,K64,D70,NK85}, clever geometrical methods to simplify networks
\cite{LF79,R80,LF82,FL88}, and the use of integral transform methods to solve
node voltages on regular networks \cite{PB55,V94,AS99,C00}.  Third, the nodes
voltages of a network through which a steady electrical current flows are
{\em harmonic} \cite{DS84}; that is, the voltage at a given node is a
suitably-weighted average of the voltages at neighboring nodes.  This same
harmonicity also occurs in the probability distribution of random walks.
Consequently, there are deep connections between the probability distribution
of random walks on a given network and the node voltages on the same network
\cite{DS84}.

Another important theme in the subject of resistor networks is the essential
role played by {\em randomness} on current-carrying properties.  When the
randomness is weak, {\em effective medium theory}
\cite{R92,B35,L52,K71,K73,K81} is appropriate to characterize how the
randomness affects the conductance.  When the randomness is strong, as
embodied by a network consisting of a random mixture of resistors and
insulators, this {\em random resistor network} undergoes a transition between
a conducting phase and an insulating phase when the resistor concentration
passes through a percolation threshold \cite{K73}.  The feature underlying 
this phase change is that for a small density of resistors, the network
consists of disconnected clusters.  However, when the resistor density passes
through the percolation threshold, a macroscopic cluster of resistors spans
the system through which current can flow.  Percolation phenomenology has
motivated theoretical developments, such as scaling, critical point
exponents, and multifractals that have advanced our understanding of
electrical conduction in random resistor networks.

This article begins with an introduction to electrical current flows in
networks.  Next, we briefly discuss analytical methods to solve the
conductance of an arbitrary resistor network.  We then turn to basic results
related to percolation: namely, the conduction properties of a large random
resistor network as the fraction of resistors is varied.  We will focus on
how the conductance of such a network vanishes as the percolation threshold
is approached from above.  Next, we investigate the more microscopic current
{\em distribution} within each resistor of a large network.  At the
percolation threshold, this distribution is {\em multifractal\/} in that all
moments of this distribution have independent scaling properties.  We will
discuss the meaning of multifractal scaling and its implications for current
flows in networks, especially the largest current in the network.  Finally,
we discuss the relation between resistor networks and random walks and show
how the classic phenomena of recurrence and transience of random walks are
simply related to the conductance of a corresponding electrical network.

The subject of current flows on resistor networks is a vast subject, with
extensive literature in physics, mathematics, and engineering journals.  This
review has the modest goal of providing an overview, from my own myopic
perspective, on some of the basic properties of random resistor networks near
the percolation threshold.  Thus many important topics are simply not
mentioned and the reference list is incomplete because of space limitations.
The reader is encouraged to consult the review articles listed in the
reference list to obtain a more complete perspective.

\section{Introduction to Current Flows}
\label{sec:intro}

In an elementary electromagnetism course, the following classic problem has
been assigned to many generations of physics and engineering students:
consider an infinite square lattice in which each bond is a 1 ohm resistor;
equivalently, the conductance of each resistor (the inverse resistance) also
equals 1.  There are perfect electrical connections at all vertices where
four resistors meet.  A current $I$ is injected at one point and the same
current $I$ is extracted at a nearest-neighbor lattice point.  What is the
electrical resistance between the input and output?  A more challenging
question is: what is the resistance between two diagonal points, or between
two arbitrary points?  As we shall discuss, the latter questions can be
solved elegantly using Fourier transform methods.

For the resistance between neighboring points, superposition provides a
simple solution.  Decompose the current source and sink into its two
constituents.  For a current source $I$, symmetry tells us that a current
$I/4$ flows from the source along each resistor joined to this input.
Similarly, for a current sink $-I$, a current $I/4$ flows into the sink along
each adjoining resistor.  For the source/sink combination, superposition
tells us that a current $I/2$ flows along the resistor directly between the
source and sink.  Since the total current is $I$, a current of $I/2$ flows
indirectly from source to sink via the rest of the lattice.  Because the
direct and indirect currents between the input and output points are the
same, the resistance of the direct resistor and the resistance of rest of the
lattice are the same, and thus both equal to 1.  Finally, since these two
elements are connected in parallel, the resistance of the infinite lattice
between the source and the sink equals 1/2 (conductance 2).  As we shall see
in Sec.~\ref{subsec:emt}, this argument is the basis for constructing an
effective medium theory for the conductance of a random network.

More generally, suppose that currents $I_i$ are injected at each node of a
lattice network (normally many of these currents are zero and there would be
both positive and negative currents in the steady state).  Let $V_i$ denote
the voltage at node $i$.  Then by Kirchhoff's law, the currents and voltages
are related by
\begin{equation}
\label{K}
I_i = \sum_j g_{ij} (V_i-V_j),
\end{equation}
where $g_{ij}$ is the conductance of link $ij$, and the sum runs over all
links $ij$.  This equation simply states that the current flowing into a node
by an external current source equals the current flowing out of the node
along the adjoining resistors.  The right-hand side of Eq.~\eqref{K} is a
{\em discrete Laplacian} operator.  Partly for this reason, Kirchhoff's law
has a natural connection to random walks.  At nodes where the external
current is zero, the node voltages in Eq.~\eqref{K} satisfy
\begin{equation}
\label{node-V}
V_i = \frac{\sum_j g_{ij}V_j}{\sum_j g_{ij}} \to \frac{1}{z}\sum_j V_j.
\end{equation}
The last step applies if all the conductances are identical; here $z$ is the
coordination number of the network.  Thus for steady current flow, the
voltage at each unforced node equals the weighted average of the voltages at
the neighboring sites.  This condition defines $V_i$ as a {\em harmonic
  function} with respect to the weight function $g_{ij}$.  

An important general question is the role of spatial disorder on current flows
in networks.  One important example is the {\em random resistor network},
where the resistors of a lattice are either present with probability $p$ or
absent with probability $1-p$ \cite{K73}.  Here the analysis tools for
regular lattice networks are no longer applicable, and one must turn to
qualitative and numerical approaches to understand the current-carrying
properties of the system.  A major goal of this article is to outline the
essential role that spatial disorder has on the current-carrying properties
of a resistor network by such approaches.

A final issue that we will discuss is the deep relation between resistor
networks and random walks \cite{DS84,L93}.  Consider a resistor network in
which the positive terminal of a battery (voltage $V=1$) is connected to a
set of boundary nodes, defined to be $\mathcal{B}_+$), and where a disjoint
set of boundary nodes $\mathcal{B}_-$ are at $V=0$.  Now suppose that a
random walk hops between nodes of the same geometrical network in which the
probability of hopping from node $i$ to node $j$ in a single step is
$g_{ij}/\sum_k g_{ik}$, where $k$ is one of the neighbors of $i$.  For this
random walk, we can ask: what is the probability $F_i$ for a walk to
eventually be absorbed on $\mathcal{B}_+$ when it starts at node $i$?  We
shall show in Sec.~\ref{sec:rwrn} that $F_i$ satisfies Eq.~\eqref{node-V}:
$F_i = \sum_j g_{ij}F_j/\sum_j g_{ij}$!  We then exploit this connection to
provide insights about random walks in terms of known results about resistor
networks and vice versa.

\section{Solving Resistor Networks}

\subsection{Fourier Transform }

The translational invariance of an infinite lattice resistor network with
identical bond conductances $g_{ij}=1$ cries out for applying Fourier
transform methods to determine node voltages.  Let's study the problem
mentioned previously: what is the voltage at any node of the network when a
unit current enters at some point?  Our discussion is specifically for the
square lattice; the extension to other lattices is straightforward.

For the square lattice, we label each site $i$ by its $x,y$ coordinates.
When a unit current is injected at $\mathbf{r_0}=(x_0,y_0)$, Eq.~\eqref{K}
becomes
\begin{equation}
\label{sq-V}
-\delta_{x,x_0}\,\,\delta_{y,y_0} = V(x+1,y)+V(x-1,y)+V(x,y+1)+V(x,y-1)-4V(x,y)\,,
\end{equation}
which clearly exposes the second difference operator of the discrete
Laplacian.  To find the node voltages, we define $V(\mathbf{k}) =
\sum_{\mathbf{r}} V(\mathbf{r})\, e^{i\mathbf{k}\cdot\mathbf{r}}~$ and then
we Fourier transform Eq.~\eqref{sq-V} to convert this infinite set of
difference equations into the single algebraic equation
\begin{equation}
\label{Vk}
V(\mathbf{k}) = \frac{e^{i\mathbf{k}\cdot\mathbf{r_0}}}{4-2(\cos k_x+\cos k_y)}~.
\end{equation}
Now we calculate $V(\mathbf{r})$ by inverting the Fourier transform
\begin{equation}
\label{Vr}
V(\mathbf{r})= \frac{1}{(2\pi)^2}\int_{-\pi}^\pi \int_{-\pi}^\pi \frac{
e^{-i\mathbf{k}\cdot(\mathbf{r}-\mathbf{r_0})}}{4-2(\cos k_x+\cos k_y)}\, \,d\mathbf{k}\,.
\end{equation}
Formally, at least, the solution is trivial.  However, the integral in the
inverse Fourier transform, known as a Watson integral \cite{W39}, is
non-trivial, but considerable understanding has gradually been developed for
evaluating this type of integral \cite{W39,PB55,V94,AS99,C00}.

For a unit input current at the origin and a unit sink of current at
$\mathbf{r_0}$, the resistance between these two points is
$V(0)-V(\mathbf{r_0})$, and Eq.~\eqref{Vr} gives
\begin{equation}
R=V(0)-V(\mathbf{r_0})=
\frac{1}{(2\pi)^2}\int_{-\pi}^\pi \int_{-\pi}^\pi \frac{
(1-e^{i\mathbf{k}\cdot\mathbf{r_0}})}
{4-2(\cos k_x+\cos k_y)}\, d\mathbf{k}\,.
\end{equation}
Tables for the values of $R$ for a set of closely-separated input and output
points are given in \cite{PB55,AS99}.  As some specific examples, for
$\mathbf{r_0}=(1,0)$, $R=\frac{1}{2}$, thus reproducing the symmetry argument
result.  For two points separated by a diagonal, $\mathbf{r_0}=(1,1)$,
$R=\frac{2}{\pi}$.  For $\mathbf{r_0}=(2,0)$, $R=2-\frac{4}{\pi}$.  Finally,
for  two points separated by a knight's move, $\mathbf{r_0}=(2,1)$,
$R=-\frac{1}{2}+\frac{4}{\pi}$.

\subsection{Direct Matrix Solution}

Another way to solve Eq.~\eqref{K}, is to recast Kirchhoff's law as the
matrix equation
\begin{equation}
\label{matrix}
I_i = \sum_{j=1}^N G_{ij} V_j,\qquad i=1,2,\ldots,N
\end{equation}
where the elements of the {\em conductance matrix} are:
\begin{eqnarray*}
G_{ij} =
\begin{cases}
 \sum _{k\ne i} g_{ik}, \quad i=j\\
     - g_{ij},\qquad i\ne j\,.
\end{cases}
\end{eqnarray*}
The conductance matrix is an example of a {\em tree matrix}, as $\mathbf{G}$
has the property that the sum of any row or any column equals zero.  An
important consequence of this tree property is that all cofactors of $G$ are
identical and are equal to the {\em spanning tree polynomial} \cite{H69}.
This polynomial is obtained by enumerating all possible tree graphs (graphs
with no closed loops) on the original electrical network that includes each
node of the network.  The weight of each spanning tree is simply the product
of the conductances for each bond in the tree.

Inverting Eq.~\eqref{matrix}, one obtains the voltage $V_i$ at each node $i$
in terms of the external currents $I_j$ ($j=1,2,\ldots,N$) and the
conductances $g_{ij}$.  Thus the two-point resistance $R_{ij}$ between two
arbitrary (not necessarily connected) nodes $i$ and $j$ is then given by
$R_{ij}=(V_i-V_j)/I$, where the network is subject to a specified external
current; for example, for the two-point geometry, $I_i=1$, $I_j=-1$, and
$I_k=0$ for $k\ne i,j$.  Formally, the two-point resistance can be written as
\cite{W82}
\begin{equation}
R_{ij} = \frac{|G^{(ij)}|}{|G^{(j)}|},
\end{equation}
where $|G^{(j)}|$ is the determinant of the conductance matrix with the
$j^{\rm th}$ row and column removed and ${|G^{(ij)}|}$ is the determinant
with the $i^{\rm th}$ and $j^{\rm th}$ rows and columns removed.  There is a
simple geometric interpretation for this conductance matrix inversion.  The
numerator is just the spanning tree polynomial for the original network,
while the denominator is the spanning tree polynomial for the network with
the additional constraint that nodes $i$ and $j$ are identified as a single
point.  This result provides a concrete prescription to compute the
conductance of an arbitrary network.  While useful for small networks, this
method is prohibitively inefficient for larger networks because the number of
spanning trees grows exponentially with network size.

\subsection{Potts Model Connection}

The matrix solution of the resistance has an alternative and elegant
formulation in terms of the spin correlation function of the $q$-state Potts
model of ferromagnetism in the $q\to 0$ limit \cite{W82,S76}.  This connection
between a statistical mechanical model in a seemingly unphysical limit and an
enumerative geometrical problem is one of the unexpected charms of
statistical physics.  Another such example is the $n$-vector model, in which
ferromagnetically interacting spins ``live'' in an $n$-dimensional spin
space.  In the limit $n\to 0$ \cite{G72}, the spin correlation functions of
this model are directly related to all self-avoiding walk configurations.

In the $q$-state Potts model, each site $i$ of a lattice is occupied by a
spin $s_i$ that can assume one of $q$ discrete values.  The Hamiltonian of
the system is
\begin{equation*}
\mathcal{H} = -\sum_{i,j} J\, \delta_{s_i,s_j},
\end{equation*}
where the sum is over all nearest-neighbor interacting spin pairs, and
$\delta_{s_i,s_j}$ is the Kronecker delta function ($\delta_{s_i,s_j}=1$ if
$s_i=s_j$ and $\delta_{s_i,s_j}=0$ otherwise).  Neighboring aligned spin
pairs have energy $-J$, while spin pairs in different states have energy
zero.  One can view the spins as pointing from the center to a vertex of a
$q$-simplex, and the interaction energy is proportional to the dot product of
two interacting spins.

The partition function of a system of $N$ spins is 
\begin{equation}
  \mathcal{Z}_N =\sum_{\{s\}} e^{\beta \sum_{i,j} J\, \delta_{s_i,s_j}}\,,
\end{equation}
where the sum is over all $2^N$ spin states ${\{s\}}$.  To make the
connection to resistor networks, notice that: (i) the exponential factor
associated with each link $ij$ in the partition function takes the values 1
or $e^{\beta J}$, and (ii) the exponential of the sum can be written as the
product
\begin{equation}
\label{Z-prod}
\mathcal{Z}_N =\sum_{\{s_i\}} \prod_{i,j} (1+v\delta_{s_i,s_j}),
\end{equation}
with $v=\tanh\beta J$.  We now make a high-temperature (small-$v$) expansion
by multiplying out the product in \eqref{Z-prod} to generate all possible
graphs on the lattice, in which each bond carries a weight
$v\delta_{s_i,s_j}$.  Summing over all states, the spins in each disjoint
cluster must be in the same state, and the last sum over the common state of
all spins leads to each cluster being weighted by a factor of $q$.  The
partition function then becomes
\begin{equation}
\mathcal{Z}_N = \sum_{\rm graphs} q^{N_c}\, v^{N_b},
\end{equation}
where $N_c$ is the number of distinct clusters and $N_b$ is the total number
of bonds in the graph.

It was shown by Kasteleyn and Fortuin \cite{KF69} that the limit $q=1$
corresponds to the percolation problem when one chooses $v=p/(1-p)$, where
$p$ is the bond occupation probability in percolation.  Even more striking
\cite{FK72}, if one chooses $v=\alpha q^{1/2}$, where $\alpha$ is a constant,
then $\lim_{q\to 0} \mathcal{Z}_N/q^{(N+1)/2} = \alpha^{N-1}T_N$, where $T_N$
is again the spanning tree polynomial; in the case where all interactions
between neighboring spins have the same strength, then the polynomial reduces
to the number of spanning trees on the lattice.  It is because of this
connection to spanning trees that the resistor network and Potts model are
intimately connected \cite{W82}.  In a similar vein, one can show that the
correlation function between two spins at nodes $i$ and $j$ in the Potts
model is simply related to the conductance between these same two nodes when
the interactions $J_{ij}$ between the spins at nodes $i$ and $j$ are equal to
the conductances $g_{ij}$ between these same two nodes in the corresponding
resistor network \cite{W82}.

\subsection{$\Delta$-Y and Y-$\Delta$ Transforms}

In elementary courses on circuit theory, one learns how to combine resistors
in series and parallel to reduce the complexity of an electrical circuit.
For two resistors with resistances $R_1$ and $R_2$ in series, the net
resistance is $R=R_1+R_2$, while for resistors in parallel, the net
resistance is $R= \left(R_1^{-1}+R_2^{-2}\right)^{-1}$.  These rules provide
the resistance of a network that contains only series and parallel
connections.  What happens if the network is more complicated?  One useful
way to simplify such a network is by the $\Delta$-Y and Y-$\Delta$ transforms
\cite{LF79,R80,LF82,FL88,SW75}.

\begin{figure}[ht]
  \begin{center}
    \includegraphics[width=0.4\textwidth]{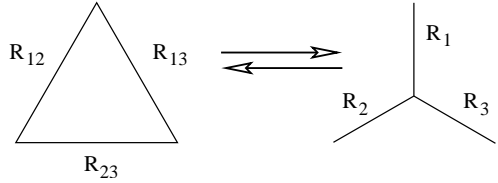}
    \caption{Illustration of the $\Delta$-Y and Y-$\Delta$ transforms.}
\label{YD}
  \end{center}
\end{figure}

The basic idea of the $\Delta$-Y transform is illustrated in Fig.~\ref{YD}.
Any triangular arrangement of resistors $R_{12}$, $R_{13}$, and $R_{23}$
within a larger circuit can be replaced by a star, with resistances $R_1$,
$R_2$, and $R_3$, such that all resistances between any two points among the
three vertices in the triangle and the star are the same.  The conditions
that all two-point resistances are the same are:
\begin{equation*}
(R_1+R_2) = \left[R_{12}^{-1}+(R_{13}+R_{23})^{-1}\right]^{-1}\equiv a_{12}
\qquad \textrm{+ cyclic permutations}.
\end{equation*}
Solving for $R_1, R_2$, and $R_3$ gives
$R_1=\frac{1}{2}(a_{12}-a_{23}+a_{13})$ + cyclic permutations; the explicit
result in terms of the $R_{ij}$ is:
\begin{equation}
\label{DY}
R_1=\frac{R_{12}R_{13}}{R_{12}+R_{13}+R_{23}} \qquad \textrm{+ cyclic permutations},
\end{equation}
as well as the companion result for the conductances $G_i=R_i^{-1}$:
\begin{equation*}
G_1=\frac{G_{12}G_{13}+G_{12}G_{23}+G_{13}G_{23}}{G_{23}}\qquad \textrm{+ cyclic permutations}.
\end{equation*}
These relations allow one to replace any triangle by a star to reduce an
electrical network.

However, sometimes we need to replace a star by a triangle to simplify a
network.  To construct the inverse Y-$\Delta$ transform, notice that the
$\Delta$-Y transform gives the resistance ratios $R_1/R_2= R_{13}/R_{23}$ +
cyclic permutations, from which $R_{13}= R_{12}(R_3/R_2)$ and
$R_{23}=R_{12}(R_3/R_1)$.  Substituting these last two results in
Eq.~\eqref{DY}, we eliminate $R_{13}$ and $R_{23}$ and thus solve for
$R_{12}$ in terms of the $R_i$:
\begin{equation}
R_{12}=\frac{R_1R_2+R_1R_3+R_2R_3}{R_3}\qquad \textrm{+ cyclic permutations},
\end{equation}
and similarly for $G_{ij}=R_{ij}^{-1}$.  To appreciate the utility of the
$\Delta$-Y and Y-$\Delta$ transforms, the reader is invited to apply them on
the Wheatstone bridge.

When employed judiciously and repeatedly, these transforms systematically
reduce planar lattice circuits to a single bond, and thus provide a powerful
approach to calculate the conductance of large networks near the percolation
threshold.  We will return to this aspect of the problem in
Sec.~\ref{subsec:exp}.

\subsection{Effective Medium Theory}
\label{subsec:emt}

Effective medium theory (EMT) determines the macroscopic conductance of a
random resistor network by a homogenization procedure
\cite{R92,B35,L52,K71,K73,K81} that is reminiscent of the Curie-Weiss
effective field theory of magnetism.  The basic idea in EMT is to replace the
random network by an effective homogeneous medium in which the conductance of
each resistor is determined self-consistently to optimally match the
conductances of the original and homogenized systems.  EMT is quite versatile
and has been applied, for example, to estimate the dielectric constant of
dielectric composites and the conductance of conducting composites.  Here
we focus on the conductance of random resistor networks, in which each
resistor (with conductance $g_0$) is present with probability $p$ and absent
with probability $1-p$.  The goal is to determine the conductance as a
function of $p$.

\begin{figure}[ht]
  \begin{center}
    \includegraphics[width=0.55\textwidth]{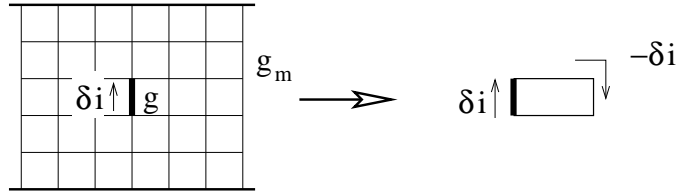}
    \caption{Illustration of EMT.  (left) The homogenized network with
      conductances $g_m$ and one bond with conductance $g$.  (right) The
      equivalent circuit to the lattice.}
\label{EMT}
  \end{center}
\end{figure}

To implement EMT, we first replace the random network by an effective
homogeneous medium in which each bond has the same conductance $g_m$
(Fig.~\ref{EMT}).  If a voltage is applied across this effective medium,
there will be a potential drop $V_m$ and a current $I_m=g_mV_m$ across each
bond.  The next step in EMT is to assign one bond in the effective medium a
conductance $g$ and adjust the external voltage to maintain a fixed total
current $I$ passing through the network.  Now an additional current $\delta
i$ passes through the conductor $g$.  Consequently, a current $-\delta i$
must flow through one terminal of $g$ to the other terminal via the remainder
of the network (Fig.~\ref{EMT}).  This current perturbation leads to an
additional voltage drop $\delta V$ across $g$.  Thus the current-voltage
relations for the marked bond and the remainder of the network are
\begin{eqnarray}
\label{emt}
I_m+\delta i &=& g(V_m+\delta V) \nonumber \\
-\delta i &=& G_{ab}\, \delta V,
\end{eqnarray}
where $G_{ab}$ is the conductance of the rest of the lattice between the
terminals of the conductor $g$.

The last step in EMT is to require that the mean value $\delta V$ averaged
over the probability distribution of individual bond conductances is zero.
Thus the effective medium ``matches'' the current-carrying properties of the
original network.  Solving Eq.~\eqref{emt} for $\delta V$, and using the
probability distribution $P(g)=p\delta(g-g_0)+(1-p)\delta(g)$ appropriate for
the random resistor network, we obtain
\begin{equation}
\label{EMT-eq}
\langle \delta V\rangle = V_m\left[\frac{(g_m-g_0)p}{(G_{ab}+g_0)}+ 
\frac{g_m(1-p)}{G_{ab}}\right]=0.
\end{equation}
It is now convenient to write $G_{ab}= \alpha g_m$, where $\alpha$ is a
lattice-dependent constant of the order of one.  With this definition,
Eq.~\eqref{EMT-eq} simplifies to 
\begin{equation}
g_m=g_o\frac{p(1+\alpha) -1}{\alpha}.
\end{equation}
The value of $\alpha$---the proportionality constant for the conductance of
the initial lattice with a single bond removed---can usually be determined by
a symmetry argument of the type presented in Sec.~\ref{sec:intro}.  For
example, for the triangular lattice (coordination number 6), the conductance
$G_{ab}=2g_m$ and $\alpha=2$.  For the hypercubic lattice in $d$ dimensions
(coordination number $z=2^d$), $G_{ab}=\frac{z-2}{2}\,g_m$.

The main features of the effective conductance $g_m$ that arises from EMT
are: (i) the conductance vanishes at a lattice-dependent percolation
threshold $p_c=1/(1+\alpha)$; for the hypercubic lattice
$\alpha=\frac{z-2}{2}$ and the percolation threshold
$p_c=\frac{2}{z}=2^{1-d}$ (fortuitously reproducing the exact percolation
threshold in two dimensions); (ii) the conductance varies linearly with $p$
and vanishes linearly in $p-p_c$ as $p$ approaches $p_c$ from above.  The
linearity of the effective conductance away from the percolation threshold
accords with numerical and experimental results.  However, EMT fails near the
percolation threshold, where large fluctuations arise that invalidate the
underlying assumptions of EMT.  In this regime, alternative methods are
needed to estimate the conductance.

\section{Conduction Near the Percolation Threshold}

\subsection{Scaling Behavior}

\begin{figure}[ht]
  \begin{center}
    \includegraphics[width=0.275\textwidth]{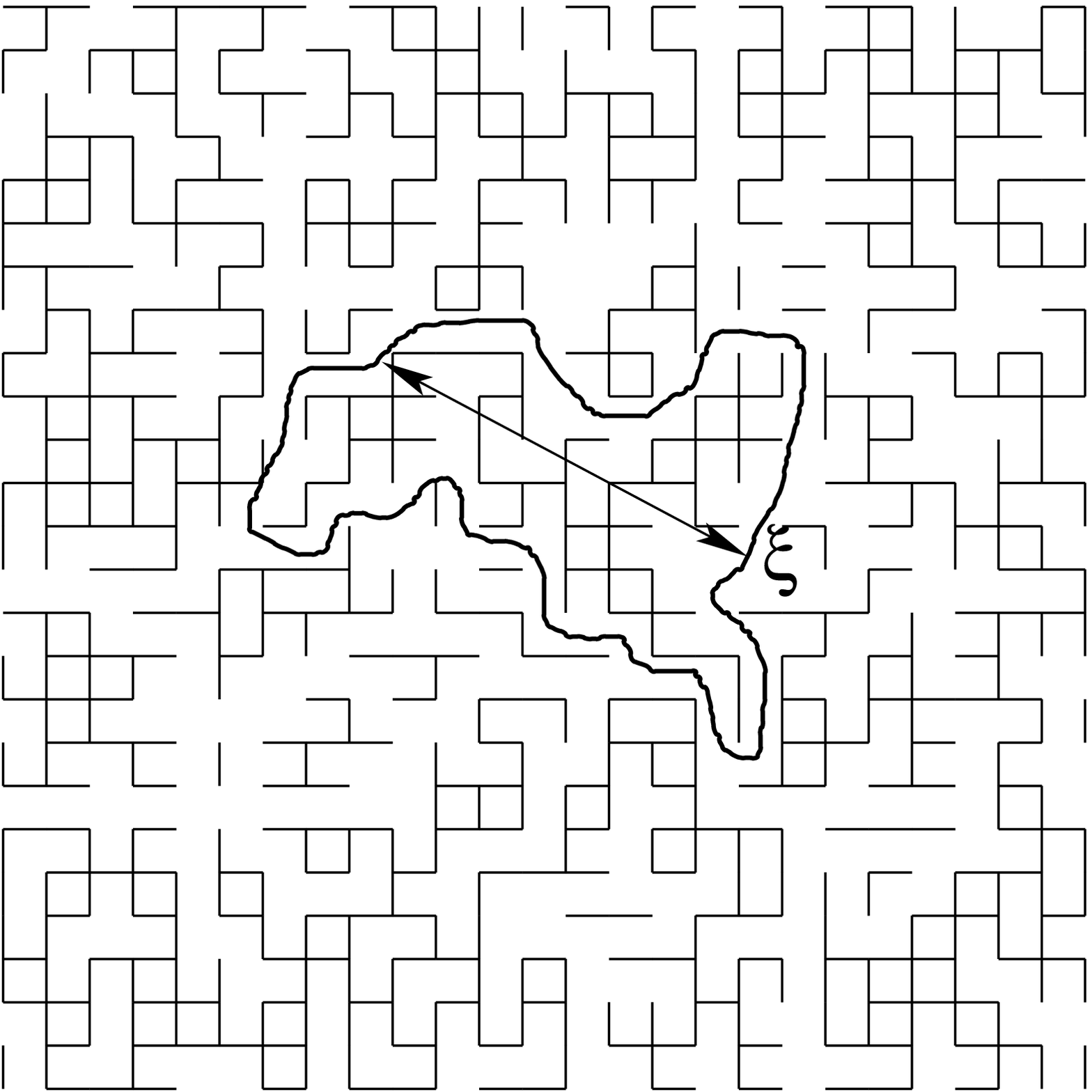}\hskip 0.8in
    \includegraphics[width=0.4\textwidth]{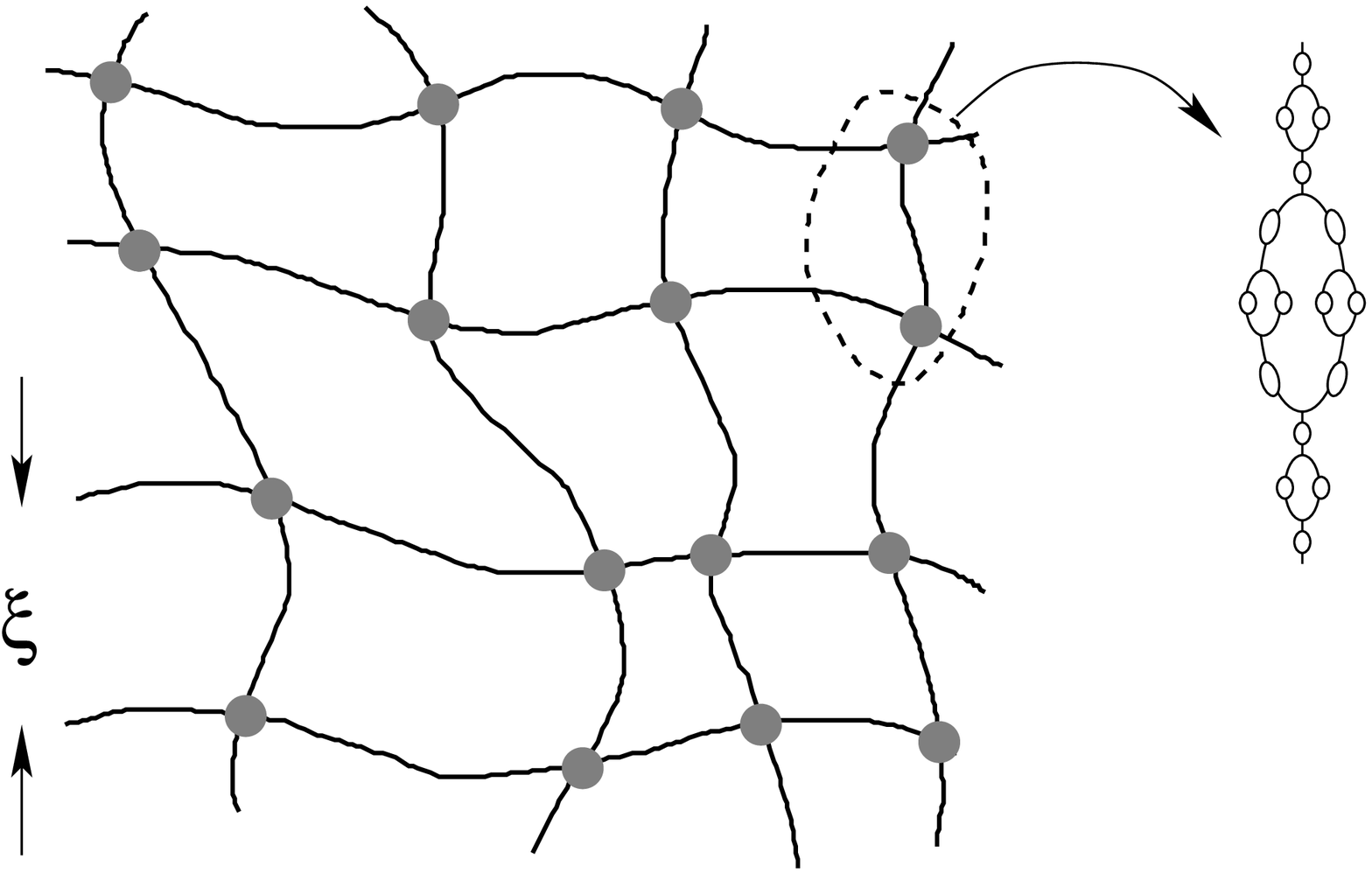}
    \caption{(left) Realization of bond percolation on a $25\times 25$ square
      lattice at p=0.505.  (Right) Schematic picture of the nodes (shaded
      circles), links and blobs picture of percolation for $p\,\,\gtwid\,\,
      p_c$.}
\label{links-blobs}
  \end{center}
\end{figure}

EMT provides a qualitative but crude picture of the current-carrying
properties of a random resistor network.  While EMT accounts for the
existence of a percolation transition, it also predicts a linear dependence
of the conductance on $p$.  However, near the percolation threshold it is
well known that the conductance varies non-linearly in $p-p_c$ near $p_c$
\cite{SA94}.  This non-linearity defines the conductance exponent $t$ by
\begin{equation}
\label{G}
G\sim (p-p_c)^t\qquad p\downarrow p_c,
\end{equation}
and much research on random resistor networks \cite{SA94} has been done to
determine this exponent.  The conductance exponent generically depends only
on the spatial dimension of the network and not on any other details (a
notable exception, however, is when link resistances are broadly distributed,
see \cite{TW85,HFN85}).  This {\em universality} is one of the central tenets
of the theory of critical phenomena \cite{S71,M76}.  For percolation, the
mechanism underlying universality is the absence of a characteristic length
scale; as illustrated in Fig.~\ref{links-blobs}, clusters on all length
scales exist when a network is close to the percolation threshold.

The scale of the largest cluster defines the correlation length $\xi$ by
$\xi\sim (p_c-p)^{-\nu}$ as $p\to p_c$.  The divergence in $\xi$ also applies
for $p>p_c$ by defining the correlation length as the typical size of finite
clusters only (Fig.~\ref{links-blobs}), thus eliminating the infinite
percolating cluster from consideration.  At the percolation threshold,
clusters on all length scales exist, and the absence of a characteristic
length implies that the singularity in the conductance should not depend on
microscopic variables.  The only parameter remaining upon which the
conductance exponent $t$ can depend upon is the spatial dimension $d$
\cite{S71,M76}.  As typifies critical phenomena, the conductance exponent has
a constant value in all spatial dimensions $d>d_c$, where $d_c$ is the upper
critical dimension which equals 6 for percolation \cite{G76b}.  Above this
critical dimension, mean-field theory (not to be confused with EMT) gives the
correct values of critical exponents.

While there does not yet exist a complete theory for the dimension dependence
of the conductance exponent below the critical dimension, a crude but useful
{\em nodes, links, and blobs} picture of the infinite cluster
\cite{SS75,G76a,S77} provides partial information.  The basic idea of this
picture is that for $p\,\gtwid\,p_c$, a large system has an irregular
network-like topology that consists of quasi-linear chains that are separated
by the correlation length $\xi$ (Fig.~\ref{links-blobs}).  For a macroscopic
sample of linear dimension $L$ with a bus bar-geometry, the percolating
cluster above $p_c$ then consists of $(L/\xi)^{d-1}$ statistically identical
chains in parallel, in which each chain consists of $L/\xi$ macrolinks in
series, and the macrolinks consists of nested blob-like structures

The conductance of a macrolink is expected to vanish as $(p-p_c)^\zeta$, with
$\zeta$ a new unknown exponent.  Although a theory for the conductance of a
single macrolink, and even a precise definition of a macrolink, is still
lacking, the nodes, links, and blobs picture provides a starting point for
understanding the dimension dependence of the conductance exponent.  Using
the rules for combining parallel and series conductances, the conductance of
a large resistor network of linear dimension $L$ is then
\begin{eqnarray}
G(p,L)\sim \left(\frac{L}{\xi}\right)^{d-1} \frac{(p-p_c)^\zeta}{L/\xi}
\sim  L^{d-2}\, (p-p_c)^{(d-2)\nu+\zeta}.
\end{eqnarray}
In the limit of large spatial dimension, we expect that a macrolink is merely
a random walk between nodes.  Since the spatial separation between nodes is
$\xi$, the number of bonds in the macrolink, and hence its resistance, scales
as $\xi^2$ \cite{S82a}.  Using the mean-field result $\xi\sim (p-p_c)^{-1/2}$,
the resistance of the macrolink scales as $(p-p_c)^{-1}$ and thus the
exponent $\zeta=1$.  Using the mean-field exponents $\nu=1/2$ and $\zeta=1$
at the upper critical dimension of $d_c=6$, we then infer the mean-field
value of the conductance exponent $t=3$ \cite{G76b,S82a,S82b}. 

Scaling also determines the conductance of a finite-size system of linear
dimension $L$ {\em exactly at\/} the percolation threshold.  Although the
correlation length formally diverges when $p-p_c=0$, $\xi$ is limited by $L$
in a finite system of linear dimension $L$.  Thus the only variable upon
which the conductance can depend is $L$ itself. Equivalently, deviations in
$p-p_c$ that are smaller than $L^{-1/\nu}$ cannot influence critical behavior
because $\xi$ can never exceed $L$.  Thus to determine the dependence of a
singular observable for a finite-size system at $p_c$, we may replace
$(p-p_c)$ by $L^{-1/\nu}$.  By this prescription, the conductance at $p_c$ of
a large finite-size system of linear dimension $L$ becomes
\begin{eqnarray}
G(p_c,L) \sim  L^{d-2} (L^{-1/\nu})^{(d-2)\nu+\zeta}\sim L^{-\zeta/\nu}~ .
\end{eqnarray}
In this {\em finite-size scaling} \cite{SA94}, we fix the occupation
probability to be exactly at $p_c$ and study the dependence of an observable
on $L$ to determine percolation exponents.  This approach provides a
convenient and more accurate method to determine the conductance exponent
compared to studying the dependence of the conductance of a large system as a
function of $p-p_c$.

\subsection{Conductance Exponent}
\label{subsec:exp}

In percolation and in the random resistor network, much effort has been
devoted to computing the exponents that characterize basic physical
observables---such as the correlation length $\xi$ and the conductance
$G$---to high precision.  There are several reasons for this focus on
exponents.  First, because of the universality hypothesis, exponents are a
meaningful quantifier of phase transitions.  Second, various observables near
a phase transition can sometimes be related by a scaling argument that leads
to a corresponding exponent relation.  Such relations may provide a
decisive test of a theory that can be checked numerically.  Finally, there is
the intellectual challenge of developing accurate numerical methods to
determine critical exponents.  The best such methods have become quite
sophisticated in their execution.

A seminal contribution was the ``theorists' experiment'' of Last and Thouless
\cite{LT71} in which they punched holes at random in a conducting sheet of
paper and measured the conductance of the sheet as a function of the area
fraction of conducting material.  They found that the conductance vanished
faster than linearly with $(p-p_c)$; here $p$ corresponds to the area
fraction of the conductor.  Until this experiment, there was a sentiment that
the conductance should be related to the fraction of material in the
percolating cluster \cite{EC70}---the percolation probability $P(p)$---a
quantity that vanished slower than linearly with $(p-p_c)$.  The reason for
this disparity is that in a resistor network, much of the percolating cluster
consists of {\em dangling ends}---bonds that carry no current---and thus make
no contribution to the conductance.  A natural geometrical quantity that
ought to be related to the conductance is the fraction of bonds $B(p)$ in the
{\em conducting backbone}---the subset of the percolating cluster without
dangling ends.  However, a clear relation between the conductivity and a
geometrical property of the backbone has not yet been established.

Analytically, there are primary two methods that have been developed to
compute the conductance exponent: the renormalization group
\cite{SW76,S78,HKL84,SKO99} and low-density series expansions
\cite{FH78,A85,AMAHK90}.  In the real-space version of the renormalization
group, the evolution of conductance distribution under length rescaling is
determined, while the momentum-space version involves a diagrammatic
implementation of this length rescaling in momentum space.  The latter is a
perturbative approach away from mean-field theory in the variable $6-d$ that
become exact as $d\to 6$.

Considerable effort has been devoted to determining the conductance exponent
by numerical and algorithmic methods.  Typically, the conductance is computed
for networks of various linear dimensions $L$ at $p=p_c$, and the conductance
exponent is extracted from the $L$ dependence of the conductance, which
should vanish as $L^{-\zeta/\nu}$.  An exact approach, but computationally
impractical for large networks, is Gauss elimination to invert the
conductance matrix \cite{SHSD83}.  A simple approximate method is Gauss
relaxation \cite{WJC75,Straley77,MAGC82,LS82,SV81} (and its more efficient
variant of Gauss-Seidel relaxation \cite{PTVF}).  This method uses
Eq.~\eqref{node-V} as the basis for an iteration scheme, in which the voltage
$V_i$ at node $i$ at the $n^{\rm th}$ update step is computed from
\eqref{node-V} using the values of $V_j$ at the $(n-1)^{\rm st}$ update in
the right-hand side of this equation.  However, one can do much better by the
conjugate gradient algorithm \cite{DBL86} and speeding up this method still
further by Fourier acceleration methods \cite{BHN86}.

Another computational approach is based on the node elimination method, in
which the $\Delta$-Y and Y-$\Delta$ transforms are used to successively
eliminate bonds from the network and ultimately reduce a large network to a
single bond \cite{R80,LF82,FL88}.  In a different vein, the transfer matrix
method has proved to be extremely accurate and efficient
\cite{DV82,Z82,DZVS84,NHH88}.  The method is based on building up the network
one bond at a time and immediately calculating the conductance of the network
after each bond addition.  This method is most useful when applied to very
long strips of transverse dimension $L$ so that a single realization gives an
accurate value for the conductance.

As a result of these investigations, as well as by series expansions for the
conductance, the following exponents have been found.  For $d=2$, where most
of the computational effort has been applied, the best estimate \cite{NHH88}
for the exponent $t$ (using $\zeta=t$ in $d=2$ only) is $t= 1.299\pm 0.002$.
One reason for the focus on two dimensions is that early estimates for $t$
were tantalizingly close to the correlation length exponent $\nu$ that is now
known to exactly equal 4/3 \cite{N79}.  Another such connection was the
Alexander-Orbach conjecture \cite{AO82}, which predicted
$t=91/72=1.2638\ldots$, but again is incompatible with the best numerical
estimate for $t$.  In $d=3$, the best available numerical estimate for $t$
appears to be $t=2.003\pm 0.047$ \cite{GL90,BT05}, while the low
concentration series method gives an equally precise result of $t=2.02\pm
0.05$ \cite{A85,AMAHK90}.  These estimates are just compatible with the
rigorous bound that $t\leq 2$ in $d=3$ \cite{G89,G90}.  In greater than three
dimensions, these series expansions give $t=2.40\pm 0.03$ for $d=4$ and
$t=2.74\pm 0.03$ for $d=5$, and the dimension dependence is consistent with
$t=3$ when $d$ reaches 6.

\section{Voltage Distribution in Random Networks}
\subsection{Multifractal Scaling}

While much research has been devoted to understanding the critical behavior
of the conductance, it was realized that the {\em distribution} of voltages
across each resistor of the network was quite rich and exhibited {\em
  multifractal\/} scaling \cite{ARC85,ARC86,RTBT85,RTT85}.  Multifractality
is a generalization of fractal scaling in which the distribution of an
observable is sufficiently broad that different moments of the distribution
scale independently.  Such multifractal scaling arises in phenomena as
diverse as turbulence \cite{M74,HP83}, localization \cite{CP86}, and
diffusion-limited aggregation \cite{HMP86,HJKPS86}.  All these diverse
examples showed scaling properties that were much richer than first
anticipated.

\begin{figure}[ht]
  \begin{center}
    \includegraphics[width=0.489\textwidth]{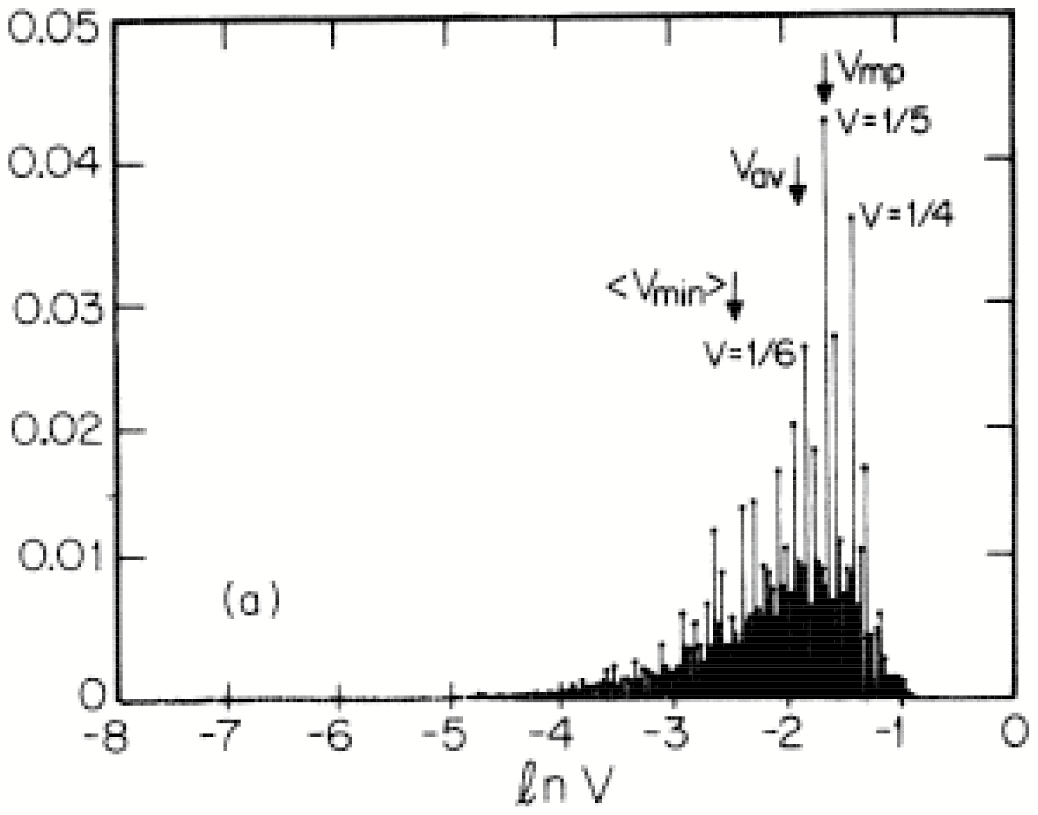}
    \includegraphics[width=0.489\textwidth]{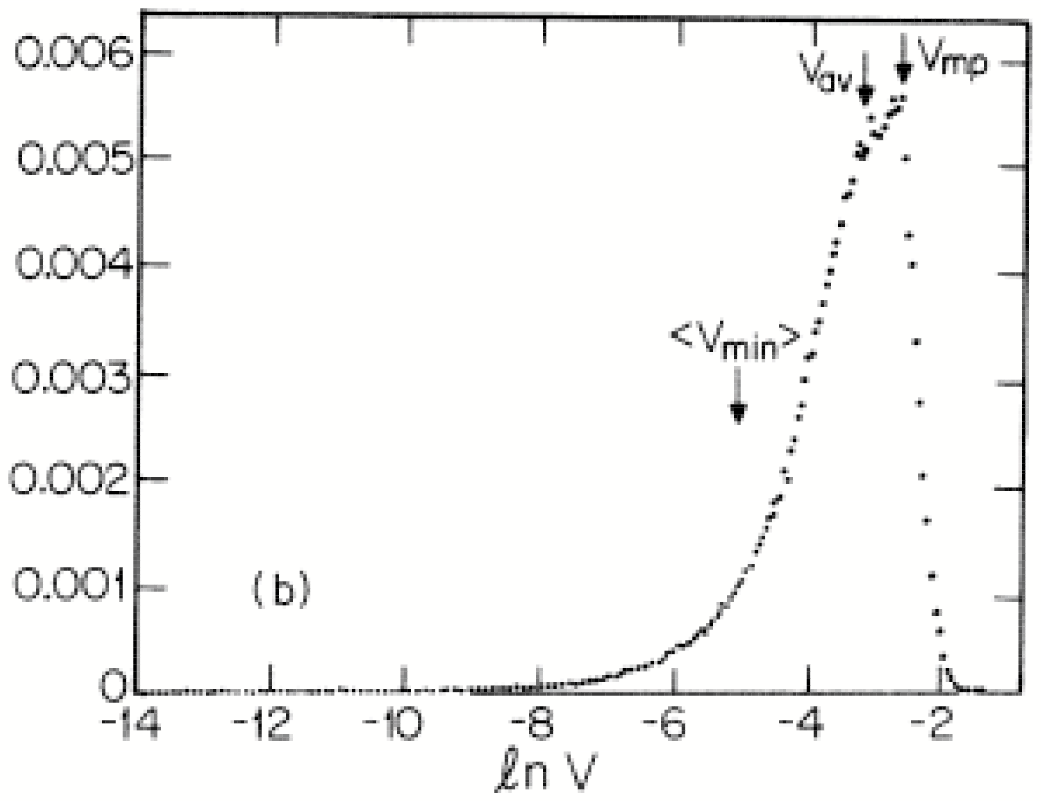}
    \includegraphics[width=0.489\textwidth]{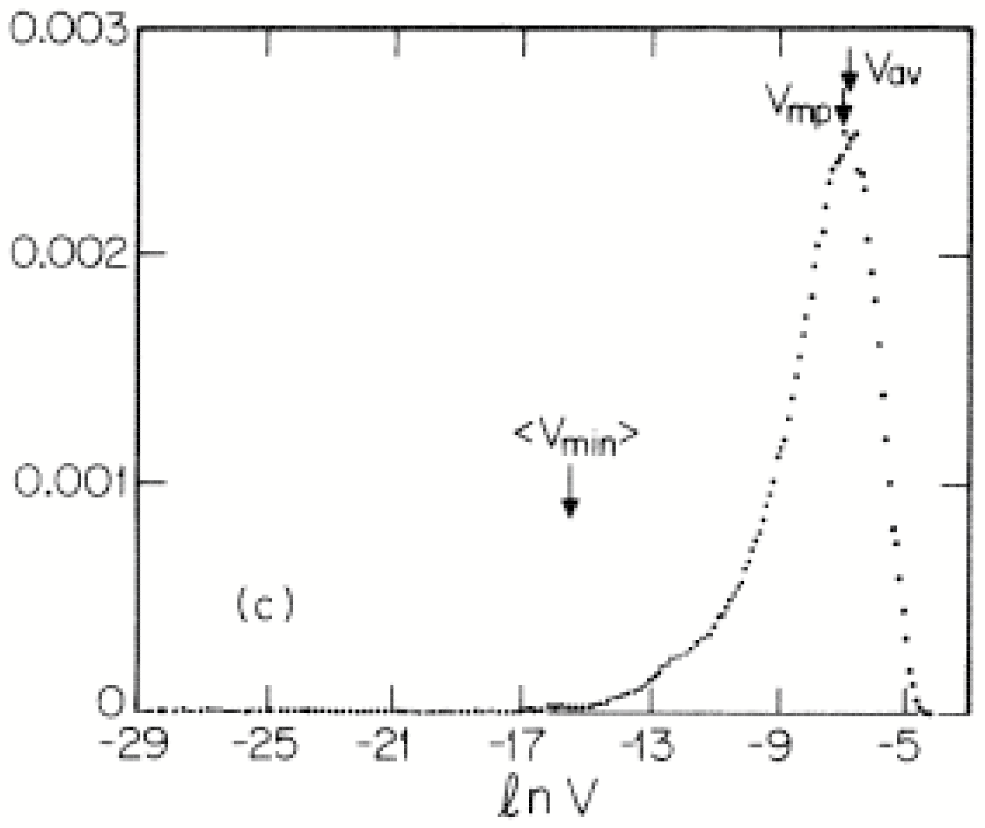}
    \caption{The voltage distribution on an $L\times L$ square lattice random
      resistor network at the percolation threshold for (a) $L=4$ (exact),
      (b) $L=10$, and (c) $L=130$.  The latter two plots are based on
      simulation data.  For $L=4$, a number of peaks, that correspond to
      simple rational fractions of the unit potential drop, are indicated.
      Also shown are the average voltage over all realizations, $V_{\rm av}$,
      the most probable voltage, $V_{\rm mp}$, and the average of the minimum
      voltage in each realization, $\langle V_{\rm min}\rangle$.  [Reprinted
      from Ref.~\cite{ARC86}].  }
\label{v-dist}
  \end{center}
\end{figure}

To make the discussion of multifractality concrete, consider the example of
the Maxwell-Boltzmann velocity distribution of a one-dimensional ideal gas
\begin{equation*}
P(v)= \sqrt{\frac{m}{2\pi k_BT}}\, e^{-mv^2/2k_BT}\equiv \frac{1}{\sqrt{2\pi
      v_{\rm th}^2}}\, e^{-v^2/2v_{\rm th}^2},
\end{equation*}
where $k_B$ is Boltzmann's constant, $m$ is the particle mass, $T$ is the
temperature, and $v_{\rm th}=\sqrt{k_BT/m}$ is the characteristic thermal
velocity.  The even integer moments of the velocity distribution are
\begin{equation*}
\langle (v^2)^n\rangle \propto (v_{\rm th}^{\,\,2})^n\equiv (v_{\rm th}^{\,\,2})^{p(n)}.
\end{equation*}
Thus a single velocity scale, $v_{\rm th}$, characterizes all positive
moments of the velocity distribution.  Alternatively, the exponent $p(n)$ is
linear in $n$.  This linear dependence of successive moment exponents
characterizes single-parameter scaling.  The new feature of multifractal
scaling is that a wide range of scales characterizes the voltage distribution
(Fig.~\ref{v-dist}).  As a consequence, the moment exponent $p(n)$ is a
non-linear function of $n$.

One motivation for studying the voltage distribution is its relation to basic
aspects of electrical conduction.  If a voltage $V=1$ is applied across a
resistor network, then the conductance $G$ and the total current flow $I$ are
equal: $I=G$.  Consider now the power dissipated through the network
$P=IV=GV^2\to G$.  We may also compute the dissipated power by adding up
these losses in each resistor to give
\begin{equation}
P=G=\sum_{ij} g_{ij} V_{ij}^2 \to \sum_{ij} V_{ij}^2=\sum_V V^2\, N(V).
\end{equation}
Here $g_{ij}=1$ is the conductance of resistor $ij$, and $V_{ij}$ is the
corresponding voltage drop across this bond.  In the last equality, $N(V)$ is
the number of resistors with a voltage drop $V$.  Thus the conductance is
just the second moment of the distribution of voltage drops across each bond in
the network.

From the statistical physics perspective it is natural to study other moments
of the voltage distribution and the voltage distribution itself.  Analogous
to the velocity distribution, we define the family of exponents $p(k)$ for
the scaling dependence of the voltage distribution at $p=p_c$ by
\begin{equation}
\label{Mk-def}
  \mathcal{M}(k)\equiv \sum_V N(V)V^k \sim L^{-p(k)/\nu}.
\end{equation}
Since $\mathcal{M}(2)$ is just the network conductance, $p(2)=\zeta$.  Other
moments of the voltage distribution also have simple interpretations.  For
example, $\langle V^4 \rangle$ is related to the magnitude of the noise in
the network \cite{RTT85,BMAH87}, while $\langle V^k \rangle$ for $k\to\infty$
weights the bonds with the highest currents, or the ``hottest'' bonds of the
network, most strongly, and they help understand the dynamics of fuse
networks of failure \cite{ARH85}.  On the other hand, negative moments weight
low-current bonds more strongly and emphasize the low-voltage tail of the
distribution.  For example, $\mathcal{M}(-1)$ characterizes hydrodynamic
dispersion \cite{KRW88}, in which passive tracer particles disperse in a
network due to a multiplicity of network paths.  In hydrodynamics dispersion,
the transit time across each bond is proportional to the inverse of the
current in the bond, while the probability for tracer to enter a bond is
proportional to the entering current.  As a result, the $k^{\rm th}$ moment
of the transit time distribution varies as $\mathcal{M}(-k+1)$, so that the
quantity that quantifies dispersion, $\langle t^2\rangle - \langle
t\rangle^2$, scales as $\mathcal{M}(-1)$.

\begin{figure}[ht]
  \begin{center}
    \includegraphics[width=0.3\textwidth]{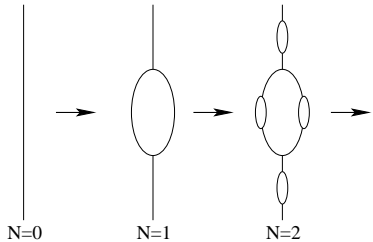}
    \caption{The first few iterations of a hierarchical model.}
\label{hierarchy}
  \end{center}
\end{figure}

A simple fractal model \cite{M82,AF89,BH91} of the conducting backbone
(Fig.~\ref{hierarchy}) illustrates the multifractal scaling of the voltage
distribution near the percolation threshold \cite{ARC85}.  To obtain the
$N^{\rm th} $-order structure, each bond in the $(N-1)^{\rm st}$ iteration is
replaced by the first-order structure.  The resulting fractal has a
hierarchical embedding of links and blobs that captures the basic geometry of
the percolating backbone.  Between successive generations, the length scale
changes by a factor of 3, while the number of bonds changes by a factor of 4.
Defining the fractal dimension $d_f$ as the scaling relation between mass
($M=4^N$) and the length scale ($\ell=3^N$) via $M\sim \ell^{d_f}$, gives a
fractal dimension $d_f=\ln 4/\ln 3$.

Now let's determine the distribution of voltage drops across the bonds.
If a unit voltage is applied at the opposite ends of a first-order structure
($N=1$) and each bond is a 1 ohm resistor, then the two resistors in the
central bubble each have a voltage drop of 1/5, while the two resistors at
the ends have a voltage drop 2/5.  In an $N^{\rm th}$-order hierarchy, the
voltage of any resistor is the product of these two factors, with number of
times each factor occurs dependent on the level of embedding of a resistor
within the blobs.  It is a simple exercise to show that the voltage
distribution is  \cite{ARC85}
\begin{equation}
N(V(j)) = 2^N \binom{N}{j}~,
\end{equation}
where the voltage $V(j)$ can take the values $2^j/ 5^N$ (with
$j=0,1,\ldots,N$).  Because $j$ varies logarithmically in $V$, the voltage
distribution is log binomial \cite{R90}.  Using this distribution in
Eq.~\eqref{Mk-def}, the moments of the voltage distribution are
\begin{equation}
\label{Mk}
\mathcal{M}(k) = \left[\frac{2(1 + 2^k)}{5^k}\right]^N.
\end{equation}
In particular, the average voltage, $\mathcal{M}(1)/\mathcal{M}(0) \equiv
V_{\rm av}$ equals $\bigl(\frac{3/2}{5}\bigr)^N$, which is very different
from the most probable voltage, $V_{\rm mp}=
\bigl(\frac{\sqrt{2}}{5}\bigr)^N$ as $N\rightarrow \infty$.  The underlying
multiplicativity of the bond voltages is the ultimate source of the large
disparity between the average and most probable values.

To calculate the moment exponent $p(k)$, we first need to relate the
iteration index $N$ to a physical length scale.  For percolation, the
appropriate relation is based on Coniglio's theorem \cite{C81}, which is a
simple but profound statement about the structure of the percolating cluster.
This theorem states that the number of singly-connected bonds in a system of
linear dimension $L$, $\mathcal{N}_s$, varies as $L^{1/\nu}$.
Singly-connected bonds are those that would disconnect the network if they
were cut.  An equivalent form of the theorem is $\mathcal{N}_s=\frac{\partial
  p'}{\partial p}$, where $p'$ is the probability that a spanning cluster
exists in the system.  This relation reflects the fact that when $p$ is
decreased slightly, $p'$ changes only if a singly-connected bond happens to
be deleted.

In the $N^{\rm th}$-order hierarchy, the number of such singly-connected
links is simply $2^N$.  Equating these two gives an effective linear
dimension, $L=2^{N\nu}$.  Using this relation in \eqref{Mk}, the moment
exponent $p(k)$ is
\begin{equation}
\label{pk}
p(k) = k-1 + \left[k\ln \left(5/4\right) - \ln (1+2^{-k})\right]/\ln 2.
\end{equation}
Because each $p(k)$ is independent, the moments of the voltage distribution
are characterized by an infinite set of exponents.  Eq.~\eqref{pk} is in
excellent agreement with numerical data for the voltage distribution in
two-dimensional random resistor networks at the percolation threshold
\cite{ARC86}.  A similar multifractal behavior was also found for the voltage
distribution of the resistor network at the percolation threshold in three
dimensions \cite{BHL96}.

\subsection*{Maximum Voltage}

An important aspect of the voltage distribution, both because of its peculiar
scaling properties \cite{DBL86} and its application to breakdown problems
\cite{ARH85,DBL86}, is the maximum voltage in a network.  The salient
features of this maximum voltage are: (i) logarithmic scaling as a function
of system size \cite{DBL86,DLB87,MG87,LD87,CMG89}, and (ii) non-monotonic
dependence on the resistor concentration $p$ \cite{KBS87}.  The former
property is a consequence of the expected size of the largest defect in the
network that gives maximal local currents.  Here, we use the terms maximum
local voltage and maximum local current interchangeably because they are
equivalent.

\begin{figure}[ht]
  \begin{center}
    \includegraphics[width=0.6\textwidth]{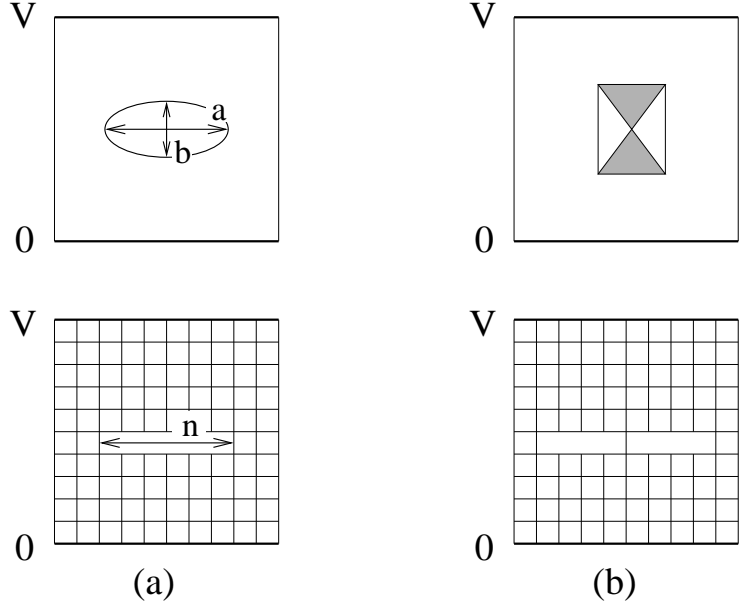}
    \caption{Defect configurations in two dimensions.  (a) An ellipse and its
      square lattice counterpart, (b) a funnel, with the region of good
      conductor shown shaded, and a 2-slit configuration on the square
      lattice. }
\label{defect}
  \end{center}
\end{figure}

To find the maximal current, we first need to identify the optimal defects
that lead to large local currents.  A natural candidate is an ellipse
\cite{DBL86,DLB87} with major and minor axes $a$ and $b$ (continuum), or its
discrete analog of a linear crack (hyperplanar crack in greater than two
dimensions) in which $n$ resistors are missing (Fig.~\ref{defect}).  Because
current has to detour around the defect, the local current at the ends of the
defect is magnified.  For the continuum problem, the current at the tip of
the ellipse is $I_{\rm tip}= I_0(1+a/b)$, where $I_0$ is the current in the
unperturbed system \cite{DBL86}.  For the maximum current in the lattice
system, one must integrate the continuum current over a one lattice spacing
and identify $a/b$ with $n$ \cite{LD87}.  This approach gives the maximal
current at the tip of a crack $I_{\rm max}\propto (1+n^{1/2})$ in two
dimensions and as $I_{\rm max}\propto (1+n^{1/2(d-1)})$ in $d$ dimensions.

Next, we need to find the size of the largest defect, which is an
extreme-value statistics exercise \cite{G58}.  For a linear crack, each
broken bond occurs with probability $1-p$, so that the probability for a
crack of length $n$ is $(1-p)^n\equiv e^{-an}$, with $a=-\ln(1-p)$.  In a
network of volume $L^d$, we estimate the size of the largest defect by $L^d
\int_{n_{\rm max}}^\infty e^{-an}\, dx=1$; that is, there exists of the order
of one defect of size $n_{\rm max}$ or larger in the network \cite{G58}.
This estimate gives $n_{\rm max}$ varying as $\ln L$.  Combining this result
with the current at the tip of a crack of length $n$, the largest current in
a system of linear dimension $L$ scales as $(\ln L)^{1/2(d-1)}$.

A more thorough analysis shows, however, that a single crack is not quite
optimal.  For a continuum two-component network with conductors of resistance
1 with probability $p$ and with resistance $r>1$ with probability $1-p$, the
configuration that maximizes the local current is a funnel \cite{MG87,CMG89}.
For a funnel of linear dimension $\ell$, the maximum current at the apex of
the funnel is proportional to $\ell^{1-\nu}$, where
$\nu=\frac{4}{\pi}\tan^{-1}(r^{-1/2})$ \cite{MG87,CMG89}.  The probability to
find a funnel of linear dimension $\ell$ now scales as $e^{-b\ell^2}$
(exponentially in its area), with $b$ a constant.  By the same extreme
statistics reasoning given above, the size of the largest funnel in a system
of linear dimension $L$ then scales as $(\ln L)^{1/2}$, and the largest
expected current correspondingly scales as $(\ln L)^{(1-\nu)/2}$.  In the
limit $r\to\infty$, where one component is an insulator, the optimal discrete
configuration in two dimensions becomes two parallel slits, each of length
$n$, between which a single resistor remains \cite{LD87}.  For this two-slit
configuration, the maximum current is proportional to $n$ in two dimensions,
rather than $n^{1/2}$ for the single crack.  Thus the maximal current 
in a system of linear dimension $L$ scales as $\ln L$ rather than as a
fractional power of $\ln L$.

The $p$ dependence of the maximum voltage is intriguing because it is
non-monotonic.  As $p$, the fraction of occupied bonds, decreases from 1,
less total current flows (for a fixed overall voltage drop) because the
conductance is decreasing, while local current in a funnel is enhanced
because such defects grow larger.  The competition between these two effects
leads to $V_{\rm max}$ attaining its peak at $p_{\rm peak}$ above the
percolation threshold that only slowly approaches $p_c$ as $L\to\infty$.  An
experimental manifestation of this non-monotonicity in $V_{\rm max}$ occurred
in a resistor-diode network \cite{RB82}, where the network reproducibly
burned (solder connections melting and smoking) when $p\simeq 0.77$, compared
to a percolation threshold of $p_c\simeq 0.58$.  Although the directionality
constraint imposed by diodes enhances funneling, similar behavior should
occur in a random resistor network.

\begin{figure}[ht]
  \begin{center}
    \includegraphics[width=0.35\textwidth]{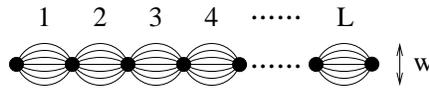}
    \caption{The bubble model: a chain of $L$ bubbles in series, each
      consisting of $w$ bonds in parallel.  Each bond is independently
      present with probability $p$.}
\label{bubble}
  \end{center}
\end{figure}

The non-monotonic $p$ dependence of $V_{\rm max}$ can be understood within
the quasi-one-dimensional ``bubble'' model \cite{KBS87} that captures the
interplay between local funneling and overall current reduction as $p$
decreases (Fig.~\ref{bubble}).  Although this system looks one-dimensional,
it can be engineered to reproduce the percolation properties of a system in
greater than one dimension by choosing the length $L$ to scale exponentially
with the width $w$.  The probability for a spanning path in this structure is
\begin{equation}
\label{pp-bubble}
p' = \left[1-(1-p)^w\right]^L \to  \exp[-L\,e^{-pw}] \qquad   L,w\to\infty,
\end{equation}
which suddenly changes from 0 to 1---indicative of percolation---at a
threshold that lies strictly within (0,1) as $L\to\infty$ and $L\sim e^w$.
In what follows, we take $L=2^w$, which gives $p_c=1/2$.  

To determine the effect of bottlenecking, we appeal to the statement of
Coniglio's theorem \cite{C81}, $\frac{\partial p'}{\partial p}$ equals the
average number of singly-connected bonds in the system.  Evaluating
$\frac{\partial p'}{\partial p}$ in Eq.~\eqref{pp-bubble} at the percolation
threshold of $p_c=\frac{1}{2}$ gives
\begin{equation}
  \frac{\partial p'}{\partial p} =w+{\cal O}(e^{-w})\sim \ln L.
\end{equation}
Thus at $p_c$ there are $w\sim \ln L$ bottlenecks.  However, current focusing
due to bottlenecks is substantially diluted because the conductance, and
hence the total current through the network, is small at $p_c$.  What is
needed is a single bottleneck of width 1.  One such bottleneck ensures the
total current flow is still substantial, while the narrowing to width 1
endures that the focusing effect of the bottleneck is maximally effective.

Clearly, a single bottleneck of width 1 occurs above the percolation
threshold.  Thus let's determine when a such an isolated bottleneck of width
1 first appears as a function of $p$.  The probability that a single
non-empty bubble contains at least two bonds is $(1-q^w-wpq^{w-1})/(1-q^w)$,
where $q=1-p$.  Then the probability $P_1(p)$ that the width of the narrowest
bottleneck has width 1 in a chain of $L$ bubbles is
\begin{eqnarray}
P_1(p) = 1-\left(1-\frac{wpq^{w-1}}{1-q^w}\right)^L
         \sim 1 - \exp\left(-Lw \frac{p}{q(1-q^w)}\, e^{-pw}\right).
\end{eqnarray}
The subtracted term is the probability that $L$ non-empty bubbles contain at
least two bonds, and then $P_1(p)$ is the complement of this quantity.  As
$p$ decreases from 1, $P_1(p)$ sharply increases from 0 to 1 when the
argument of the outer exponential becomes of the order of 1; this change
occurs at $\hat p \sim p_c + {\cal O}(\ln(\ln L)/\ln L)$.  At this point, a
bottleneck of width 1 first appears and therefore $V_{\rm max}$ also occurs
for this value of $p$.

\section{Random Walks and Resistor Networks}
\label{sec:rwrn}

\subsection{The Basic Relation}

\begin{figure}[ht]
  \begin{center}
    \includegraphics[width=0.6\textwidth]{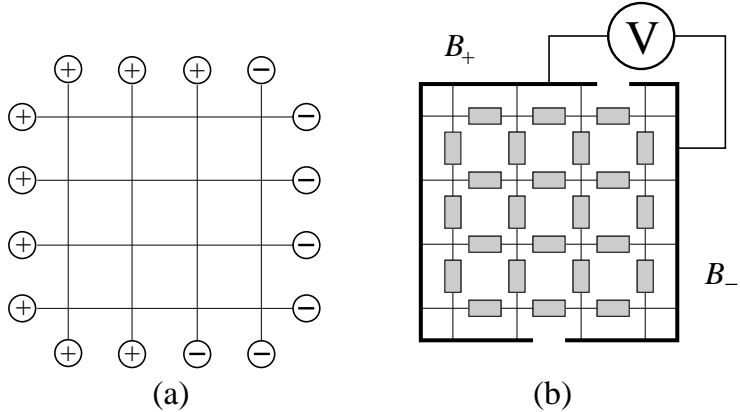}
    \caption{(a) Lattice network with boundary sites $\mathcal{B}_+$ or
      $\mathcal{B}_-$.  (b) Corresponding resistor network in which each
      rectangle is a 1 ohm resistor.  The sites in $\mathcal{B}_+$ are all
      fixed at potential $V=1$, and sites in $\mathcal{B}_-$ are all
      grounded.}~\label{network}
  \end{center}
\end{figure}

We now discuss how the voltages at each node in a resistor network and the
resistance of the network are directly related to {\em first-passage}
properties of random walks \cite{F68,L93,W94,R01}.  To develop this
connection, consider a random walk on a finite network that can hop between
nearest-neighbor sites $i$ to $j$ with probability $p_{ij}$ in a single step.
We divide the boundary points of the network into two disjoint classes,
$\mathcal{B}_+$ and $\mathcal{B}_-$, that we are free to choose; a typical
situation is the geometry shown in Fig.~\ref{network}.  We now ask: starting at
an arbitrary point $i$, what is the probability that the walk {\em
  eventually} reaches the boundary set $\mathcal{B}_+$ without first reaching
any node in $\mathcal{B}_-$?  This quantity is termed the exit probability
$\mathcal{E}_+(i)$ (with an analogous definition for the exit probability
$\mathcal{E}_-(i)=1- \mathcal{E}_+(i)$ to $\mathcal{B}_-$).

We obtain the exit probability $\CE_+(i)$ by summing the probabilities for
all walk trajectories that start at $i$ and reach a site in
$\mathcal{B}_+$ without touching any site in $\mathcal{B}_-$ (and similarly
for $\CE_-(i)$).  Thus
\begin{equation}
  \label{c1:poisson-exit-rec}
\CE_\pm(i)=\sum_{p_\pm} \CP_{p_\pm}(i),
\end{equation}
where $\CP_{p_\pm}(i)$ denotes the probability of a path from $i$ to
$\mathcal{B}_\pm$ that avoids $\mathcal{B}_\mp$.  The sum over all these
restricted paths can be decomposed into the outcome after one step, when the
walk reaches some intermediate site $j$, and the sum over all path remainders
from $j$ to $\mathcal{B}_\pm$.  This decomposition gives
\begin{eqnarray}
  \label{c1:poisson-exit-rec-1}
\CE_\pm(i) &=& \sum_j p_{ij}\, \CE_\pm(j).
\end{eqnarray}
Thus $\CE_\pm(i)$ is a harmonic function because it equals a weighted average
of $\CE_\pm$ at neighboring points, with weighting function $p_{ij}$.  This
is exactly the same relation obeyed by the node voltages in
Eq.~\eqref{node-V} for the corresponding resistor network when we identify
the single-step hopping probabilities $p_{ij}$ with the conductances
$g_{ij}/\sum_j g_{ij}$.  We thus have the following equivalence:
\begin{itemize}
\item Let the boundary sets $\mathcal{B}_+$ and $\mathcal{B}_-$ in a resistor
  network be fixed at voltages 1 and 0 respectively, with $g_{ij}$ the
  conductance of the bond between sites $i$ and $j$.  Then the voltage at any
  interior site $i$ coincides with the probability for a random walk, which
  starts at $i$, to reach $\mathcal{B}_+$ before reaching $\mathcal{B}_-$,
  when the hopping probability from $i$ to $j$ is $p_{ij}=g_{ij}/\sum_j
  g_{ij}$.
\end{itemize}

If all the bond conductances are the same---corresponding to single-step
hopping probabilities in the equivalent random walk being identical---then
Eq.~\eqref{c1:poisson-exit-rec-1} is just the discrete Laplace equation.  We
can then exploit this correspondence between conductances and hopping
probabilities to infer non-trivial results about random walks and about
resistor networks from basic electrostatics.  This correspondence can also be
extended in a natural way to general random walks with a spatially-varying
bias and diffusion coefficient, and to continuous media.

The consequences of this equivalence between random walks and resistor
networks is profound.  As an example \cite{R01}, consider a diffusing
particle that is initially at distance $r_0$ from the center of a sphere of
radius $a<r_0$ in otherwise empty $d$-dimensional space.  By the
correspondence with electrostatics, the probability that this
particle eventually hits the sphere is simply the electrostatic potential at
$r_0$, $\CE_-(r_0)=(a/r_0)^{d-2}$\,!

\subsection{Network Resistance and P\'olya's Theorem}

An important extension of the relation between exit probability and node
voltages is to infinite resistor networks.  This extension provides a simple
connection between the classic recurrence/transience transition of random
walks on a given network \cite{F68,L93,W94,R01} and the electrical resistance
of this same network \cite{DS84}.  Consider a symmetric random walk on a
regular lattice in $d$ spatial dimensions.  Suppose that the walk starts at
the origin at $t=0$.  What is the probability that the walk {\em eventually}
returns to its starting point?  The answer is strikingly simple:
\begin{itemize}
\item For $d\leq 2$, a random walk is {\em certain} to eventually return to
  the origin.  This property is known as {\em recurrence}.
\item For $d>2$, there is a non-zero probability that the random walk will
  {\em never} return to the origin.  This property is known as {\em
    transience}.
\end{itemize}

Let's now derive the transience and recurrence properties of random walks in
terms of the equivalent resistor network problem.  Suppose that the voltage
$V$ at the boundary sites $\mathcal{B}_+$ is set to one.  Then by Kirchhoff's
law, the total current entering the network is
\begin{eqnarray}
\label{c1:I-tot}
I = \sum_j (1-V_j)g_{+j}= \sum_j (1-V_j)p_{+j}\sum_k g_{+k}.
\end{eqnarray}
Here $g_{+j}$ is the conductance of the resistor between $\mathcal{B}_+$ and
a neighboring site $j$, and $p_{+j}=g_{+j}/\sum_j g_{+j}$.  Because the
voltage $V_j$ also equals the probability for the corresponding random walk
to reach $\mathcal{B}_+$ without reaching $\mathcal{B}_-$, the term
$V_j\,p_{+j}$ is just the probability that a random walk starts at
$\mathcal{B}_+$, makes a single step to one of the sites $j$ adjacent to
$\mathcal{B}_+$ (with hopping probability $p_{ij}$), and then returns to
$\mathcal{B}_+$ without reaching $\mathcal{B}_-$.  We therefore deduce that
\begin{eqnarray}
\label{c1:I-tot-p}
I = \sum_j  (1  -V_j)g_{+j}
 &=& \sum_k g_{+k}\sum_j (1  -V_j)p_{+j}\nonumber\\
    &=& \sum_k g_{+k}\times (1-{\rm return\ probability}) \nonumber\\ 
    &=& \sum_k g_{+k}\times{\rm escape\ probability}.
\end{eqnarray}
Here ``escape'' means that the random walk reaches the set $\mathcal{B}_-$
without returning to a node in $\mathcal{B}_+$.

\begin{figure}[ht]
  \begin{center}
    \includegraphics[width=0.25\textwidth]{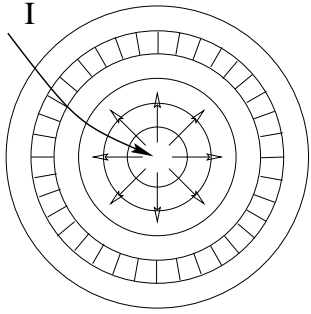}
 \caption{Decomposition of a conducting medium into concentric shells, each
   of which consists of fixed-conductance blocks.  A current $I$ is injected
   at the origin and flows radially outward through the medium.}~\label{shell}
  \end{center}
\end{figure}

On the other hand, the current and the voltage drop across the network are
related to the conductance $G$ between the two boundary sets by $I=GV=G$.
From this fact, Eq.~(\ref{c1:I-tot-p}) gives the fundamental result
\begin{equation}
\label{c1:p-esc}
{\rm escape\ probability} \equiv P_{\rm escape} = \frac{G}{\sum_k g_{+k}}.
\end{equation}
Suppose now that a current $I$ is injected at a single point of an infinite
network, with outflow at infinity (Fig.~\ref{shell}).  Thus the probability
for a random walk to never return to its starting point, is simply
proportional to the conductance $G$ from this starting point to infinity of
the same network.  Thus a subtle feature of random walks, namely, the escape
probability, is directly related to currents and voltages in an equivalent
resistor network.

Part of the reason why this connection is so useful is that the conductance
of the infinite network for various spatial dimensions can be easily
determined, while a direct calculation of the return probability for a random
walk is more difficult.  In one dimension, the conductance of an infinitely
long chain of identical resistors is clearly zero.  Thus $P_{\rm escape}=0$
or, equivalently, $P_{\rm return}=1$.  Thus a random walk in one dimension is
recurrent.  As alluded to at the outset of Sec.~\ref{sec:intro}, the
conductance between one point and infinity in an infinite resistor lattice in
general spatial dimension is somewhat challenging.  However, to merely
determine the recurrence or transience of a random walk, we only need to know
if the return probability is zero or greater than zero.  Such a simple
question can be answered by a crude physical estimate of the network
conductance.

To estimate the conductance from one point to infinity, we replace the
discrete lattice by a continuum medium of constant conductance.  We then
estimate the conductance of the infinite medium by decomposing it into a
series of concentric shells of fixed thickness $dr$.  A shell at radius $r$
can be regarded as a parallel array of $r^{d-1}$ volume elements, each of
which has a fixed conductance.  The conductance of one such shell is
proportional to its surface area, and the overall resistance is the sum of
these shell resistances.  This reasoning gives
\begin{eqnarray}
\label{c1:R-inf}
R \sim\int^\infty R_{\rm shell}(r)\,dr &\sim&  \int^\infty
\frac{dr}{r^{d-1}}
  =\begin{cases}\infty & \text{for}\ d\leq 2\\ \\ (P_{\rm escape}\sum_j g_{+j})^{-1} & 
\text{for}\ d>2
\end{cases}
\end{eqnarray}

The above estimate gives an easy solution to the recurrence/transience
transition of random walks.  For $d\leq 2$, the conductance to infinity is
zero because there are an insufficient number of independent paths from the
origin to infinity.  Correspondingly, the escape probability is zero and the
random walk is recurrent.  The case $d=2$ is more delicate because the
integral in Eq.~\eqref{c1:R-inf} diverges only logarithmically at the upper
limit.  Nevertheless, the conductance to infinity is still zero and the
corresponding random walk is recurrent (but just barely).  For $d>2$, the
conductance between a single point and infinity in an infinite homogeneous
resistor network is non zero and therefore the escape probability of the
corresponding random walk is also non zero---the walk is now transient.

There are many amusing ramifications of the recurrence of random walks and we
mention two such properties.  First, for $d\leq 2$, even though a random walk
eventually returns to its starting point, the mean time for this event is
infinite!  This divergence stems from a power-law tail in the time dependence
of the first-passage probability \cite{R01,F68}, namely, the probability that
a random walk returns to the origin for the first time.  Another striking
aspect of recurrence is that because a random walk returns to its starting
point with certainty, it necessarily returns an infinite number of times.

\section{Future Directions}

There is a good general understanding of the conductance of resistor
networks, both far from the percolation threshold, where effective medium
theory applies, and close to percolation, where the conductance $G$ vanishes
as $(p-p_c)^t$.  Many advancements in numerical techniques have been
developed to determine the conductance accurately and thereby obtain precise
values for the conductance exponent, especially in two dimensions.  In spite
of this progress, we still do not yet have the right way, if it exists at
all, to link the geometry of the percolation cluster or the conducting
backbone to the conductivity itself.  Furthermore, many exponents of
two-dimensional percolation are known exactly.  Is it possible that the exact
approaches developed to determine percolation exponents can be extended to
give the exact conductance exponent?

Finally, there are aspects about conduction in random networks that are worth
highlighting.  The first falls under the rubric of {\em directed percolation}
\cite{DP}.  Here each link in a network has an intrinsic directionality that
allows current to flow in one direction only---a resistor and diode in
series.  Links are also globally oriented; on the square lattice for example,
current can flow rightward and upward.  A qualitative understanding of
directed percolation and directed conduction has been achieved that parallels
that of isotropic percolation.  However, there is one facet of directed
conduction that is barely explored.  Namely, the state of the network (the
bonds that are forward biased) must be determined self consistently from the
current flows.  This type of non-linearity is much more serious when the
circuit elements are randomly oriented.  These questions about the coupling
between the state of the network and its conductance are central when the
circuit elements are intrinsically non-linear \cite{KS82,RH87}.  This is a
topic that seems ripe for new developments.

\end{document}